# OBSERVATIONS OF SERVICE IDENTIFICATION FROM TWO ENTERPRISES

Ville Alkkiomäki and Kari Smolander

Lappeenranta University of Technology,
P.O. Box 20, 53851 Lappeenranta, Finland, EU

## ABSTRACT

*Service-oriented computing has created new requirements for information systems development processes and methods. The adoption of service-oriented development requires service identification methods matching the challenge in enterprises. A wide variety of service identification methods (SIM) have been proposed, but less attention has been paid to the actual requirements of the methods. This paper provides an ethnographical look at challenges in service identification based on data from 14 service identification sessions, providing insight into the practice of service identification. The findings identified two types of service identification sessions and the results can be used for selecting the appropriate SIM based on the type of the session.*

## 1. INTRODUCTION

Service identification is one of the most challenging aspects of service-oriented architecture (SOA) and is also a concrete example of business/IT communication, requiring methods that address the key elements of SOA: services, service compositions, flows of information and the implementation of the services [1].

Information systems (IS) development methods have been a focus area in systems engineering research. A wide variety of methods have been proposed for IS development with the promise of streamlined and more predictable way of working. In spite of this extensive work, the feedback from the industry on predefined methods has not been completely positive [2, 3]. Generic context-independent methods do not seem to solve the problems of practitioners and those systematic methods that enterprises reported to use in their service identification are often self-developed [4, 5].

The actual needs from the enterprise context, what challenges the service identification method (SIM) should solve, are not widely studied, and the challenges of SOA adoption in general has been studied mainly with surveys and case studies [6-10]. This paper provides a novel view to the challenges of service identification in enterprises using an ethnographical approach. By using ethnography, we will gain first-hand observations and evidence of service identification in practice.

The research questions of the paper are:

RQ1: What are the requirements for service identification method to facilitate business/IT communication?

RQ2: How does the use of an SIM affect the service elicitation sessions?

The paper is structured as follows: the next section provides a literature review of the service identification methods for enterprises, which will be followed by a section describing the research process of this paper. Section 4 publishes the results of the study, followed by discussion and conclusion.

## 2. SERVICE IDENTIFICATION METHODS

The idea behind SOA is to package software resources as reusable and autonomous services representing business-complete work that can be used in business processes as flexible building blocks for business development [11]. Another aspect of SOA, differing fundamentally from conventional software development, is the dynamical access to services at runtime. These aspects require new methods to design the services and their compositions at an appropriate level of granularity [1].

A variety of service identification methods has been developed over the years, and based on a systematic literature review classifying 105 methods (2002 to June 2013), the methods can be categorized based on their strategy into three categories: top-down, bottom-up and meet-in-the-middle [12]. SIMs with bottom-up strategy attempts to provide an inventory analysis prior to the design and development of the services, top-down strategy focuses on the fulfilment of the immediate business requirements and the meet-in-the-middle is a combination of both [12].

Service identification is considered to be one of the most practical phases of SOA and meet-in-the-middle methods, being the most complete approach as it addresses both the business and the technical perspectives [12]. Still, only 12 SIMs out of the 105 methods identified by systematic literature review, used this meet-in-the-middle approach [12]. Furthermore, only 2 [13, 14] out of these 12 meet-in-the-middle methods have been published with real experiments from enterprises. Furthermore, the business aspect of the service-oriented design has been neglected in enterprises as well, considering SOA as a technical solution to wrap existing software assets and focusing the service identification efforts on the fine-grained services only [15].

Taking a service-oriented architecture into use as such does not deliver reusable services or business agility. Many factors have been recognized as SOA enablers, including better facilitation of IT/business communication [6-8], top management support [10], use of standards [8] and SOA competence [8]. The main common obstacles mentioned include lack of business involvement in SOA development [6, 9] and change-resistant personnel [9].

Ramasubbu and Balan [3] report how the development process selections were made in 112 software projects conducted in high maturity companies (CMM level 5). The main take-away of the survey was that the process choice matters and that the projects adopting agile methods even in a high maturity environment, performed better than the ones using standard plan-driven processes [3]. It is also notable that practitioners were more skeptical towards IS development methods than their managers [16] and that the more hierarchical the organization, the higher the adoption rate of rigid software development methods was [17].

Several comparative evaluations of the existing service identification methods have been made, providing generic guidance for enterprises on the availability of SIMs and their suitability to specific needs. Gu and Lago [18] analyzed 12 selected SIMs based on their feature coverage over the lifecycle of service development. They also provide selection guidelines covering several aspects of SIMs, but leave the selection of right criteria and the method for enterprises. Gholami et al. [19] provide a similar framework for evaluating features of SIMs, but without further guidance in the enterprise context. Similarly Ramollari et al. [20] provide feature comparison of 10 SIMs, and Kontogogos and Avgeriou make another comparison of 7 SIMs [21]. None of the above comparisons make any statement of what the actual requirements or needs of enterprises are for an SIM.

The developed methods can be seen as one form of IT artifacts [22]. However, McKay et al. [23] argues that the conceptualization of the IT artifact is too narrow, if it includes only the constructs, models and methods, but excludes the surrounding people and the organization as proposed by Hevner et al. [22]. Instead of scoping out the surrounding context, McKay et al. proposes the socio-technical view and a richer construction-centered perspective taking into account, not only the IT artifacts, but also their effect on the management system of the organization and human activity, and how the IT artifacts appear to the user [23]. Also, Sein et al. [24] argue that design research does not fully recognize the role of the organizational context when used in IS development and proposes to fill in the gaps by cross-fertilizing the design research with action research. The importance of the context for building the system has been long recognized in the field of requirements engineering [25].

The current research is lacking empirical testing of service identification methods in enterprises, especially for SIMs using meet-in-the-middle strategy requiring involvement of different kinds of stakeholders. This paper provides insights into the service identification needs of enterprises by observing and analyzing the practice and the challenges of 14 service identification sessions directly.

## 3. RESEARCH DESIGN

Ethnography is an in-depth method used to study the culture and organizational structure related to information systems [26]. The core of this approach is to study people for a long period of time in their own environment and to produce descriptive data free from imposed external concepts and ideas [26].

To answer the research questions, participant observation in two enterprises and a set of real world service identification sessions were documented using a predefined session report template. These documented sessions include both sessions orchestrated with and without a service identification method. The reports were documented either during the session or immediately

International Journal of Software Engineering & Applications (IJSEA), Vol.6, No.2, March 2015afterwards. To keep the sessions as authentic as possible and to avoid any additional tension that could disturb the actual work, these sessions were not audio recorded and the participants were informed retrospective of the research. The research purpose was, explained to them prior to the submission of this paper, and their consent for participation in the research was complied with.

The participants of the studied service identification sessions were categorized into business and technical stakeholders based on their role in the session. The term business stakeholder is used to refer to a person who is defining the end user needs along with the business processes and the business justification. Business stakeholders are assumed to know what the application should do and how the changes affect the business. Typical positions of the business stakeholders were program or project managers and directors or managers of a certain business domain.

Similarly the term technical stakeholder is used to refer to a person who is designing the identified services with an adequate understanding of the technical constraints and/or availability of the existing IT assets. Typical positions of the technical stakeholders were technical experts, account managers of external suppliers, and IT architects.

To compare the discussions of the service identification sessions, it became clear after first three sessions, that a common reference was needed. The Zachman framework [27] was chosen as the reference to classify the discussions during the sessions 4-15. The Zachman framework [27] providing ontology of topics or "cells" relevant for enterprises, see Table 1. We considered it to be generic enough as a reference to document and evaluate the discussion topics. Additionally, any topics identified outside the Zachman framework were recorded as memos. To make the material comparable, the Zachman classification for the first 3 sessions was reconstructed afterwards based on the written memos.

Table 1. Cells of the Zachman framework based on [27]

|  | Data | Function | Network | People | Time | Motivation |
|---|---|---|---|---|---|---|
| Contextual | List of things | List of processes | List of locations | List of organizations | List of events | List of goals |
| Conceptual | Semantic model | Business process model | Logistics network | Work flow model | Master schedule | Business plan |
| Logical | Logical data model | Application architecture | Distributed system architecture | Human interface architecture | Processing structure | Business rule model |
| Physical | Physical data model | System design | System architecture | Presentation architecture | Control structure | Rule design |
| Detailed | Data definition | Program | Network architecture | Security architecture | Timing definition | Rule specification |

Integration Use Case (IUC) method [28] was used to some extent in the company A and was chosen as a service identification method to be used in sessions. The idea behind IUC is to make explicit the services between systems in a complex enterprise system landscape [28]. Another target is to have an architectural view of the project that can be checked against business process specifications [28]. The IUC technique itself does not provide guidelines of how to use it to facilitate service identification sessions, but it was simple enough to be explained and taken into use ad-hoc during the sessions.

An Integration Use Case (IUC) represents the abstract service interface between a service provider(s) and a service consumer(s). If an enterprise service bus (ESB), a messaging queue, or another middleware system is used in system integration, then the integration use case also describes the role and actions of the middleware between the systems as shown in Figure 1 below.

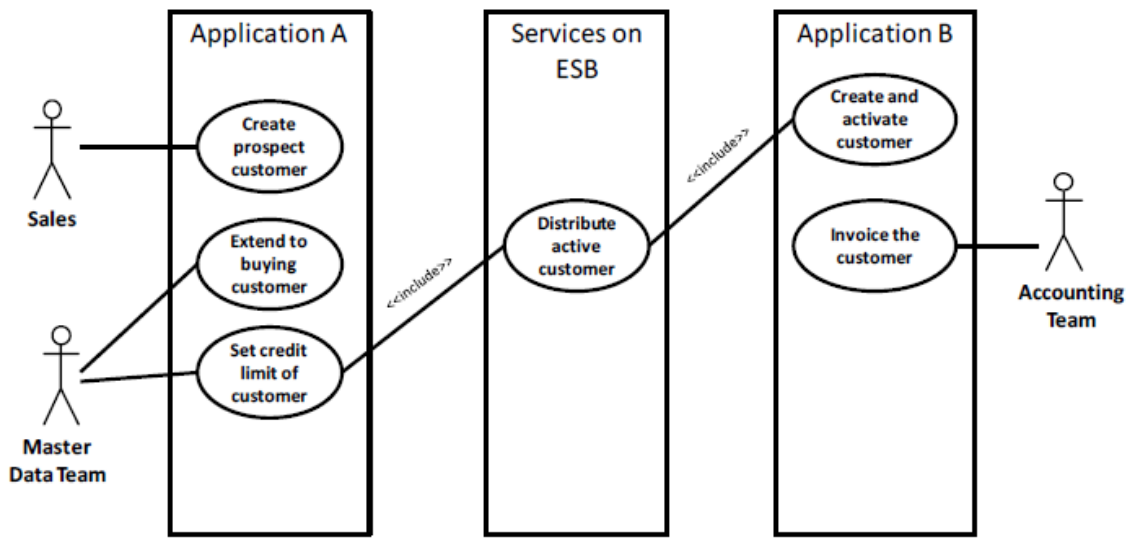

Figure 1. Example of Integration Use Case (IUC) diagram

## 4. RESULTS

Total of 14 service identification sessions were observed and analyzed, with a total of 40 stakeholders participating in one or more sessions. The data was gathered from two companies during 2009-2013.

The company A provides logistics and e-commerce services, has a net sales of roughly 2 B€ and employs over 20000 people in over 10 countries. The company B has global operations in heavy manufacturing industry and services, has a net sales of roughly 5B€ and employs roughly 20000 persons.

An example of the observations gathered from the sessions is provided in Table 2 below. The Zachman framework was used as a reference to describe the abstraction level and topics of the discussions with and without the IUC method.

Table 2. Example how service identification session report

| Session Topic (stakeholder type) | Discussion topics without IUC (X) | Discussion topics with IUC (O) |
|---|---|---|
| Project X service identification meeting | X X X X X X | O O    O O |
| Participants: |  X      X | O O O O O O |
| Person A (business stakeholder) | | O O O   O |
| Person B (business stakeholder) | | |
| Person C (technical stakeholder) | | |
| Person D (business stakeholder) | | |

The IUC approach was explained and taken into use in four of the sessions, providing a way evaluate the change in the discussion topics with and without using a SIM during the same session. Once the IUCs were introduced to the stakeholders, some of the stakeholders used the same IUCs as the basis for the follow-up sessions. In total, the IUC method was used in some form in eight sessions out of fourteen. Respectively, SIM was not used in six sessions.

As the analyzed sessions had differences, the requirements for SIM also differed. Some of the sessions clearly didn't require any SIM, as the requirements and constraints were clear enough to be documented, and there were no conflicts between the stakeholders to resolve. Still, the IUCs were used in some of these sessions to document the functional requirements of the services with the business stakeholders as well as to explain the requirements to the technical stakeholders. The notation was well understood especially by the technical stakeholders and the documentation produced by the method was detailed enough to identify conflicts with known technical constraints.

Sessions with conflicting requirements and/or technical constraints limiting the solution space benefited from the use of the SIM, which helped to specify conflicting topics and suggest alternative solutions for a further discussion. In general, IUCs seemed a practical way to facilitate service identification sessions, and participants needed only a little training to use them. However, video conferencing limited the possibilities of using IUCs as a means to facilitate the discussion, as it was much slower to draw them during the session using a modeling tool than with a white board.

Another facilitation method was the use of snapshots of the application's user interface, which was also found to be a practical way to facilitate the discussion. However, with this approach, the services were discussed only indirectly and required additional knowledge of the used applications to translate the discussion between the stakeholders using two or more different application terminologies. Another disadvantage of using application user interfaces to facilitate the sessions was the tendency to focus on too detailed topics.

In summary, the reported sessions can be categorized into functional service requirement documentation and business/SOA alignment sessions. This categorizing is shown in Table 3 below with the characteristics of the categories, the identified requirements for a SIM, and the number of sessions where each Zachman cell was discussed with and without an SIM.

As a response to RQ1, Table 3 shows that the requirements for SIM differed depending on the kind of session. The common requirement (R1) was to provide a systematic approach to facilitate the session and its discussion. For sessions requiring business/SOA alignment, the SIM should provide (R2) a common ground for both technical and business stakeholders, using concepts that can be linked both with the business processes, as well as with the potential technical implementations.

Table 3 also provides an overview of the results relevant to RQ2. The use of an SIM clearly changed the discussion from the conceptual/contextual layer to the contextual/logical level. The use of UI screenshots to facilitate a session resulted in a change of discussion to physical/detailed layers.

Furthermore, the abstraction level of the discussion became more concrete in the few sessions where IUCs were taken into use during the session itself. The reason for this might have been that, with the IUCs drawn on a white board, there was a common visual reference to facilitate the discussion and to make more specific questions on individual services, for example "in which phase of the process does system X send the sales order message to system Y?" or "does system X know the customer's id in this phase of the process?" From a technical person's point of view, the functional requirements of the services were concrete enough to comment on how well they matched with existing services and whether new service aggregates needed to be created to fulfill the requirements.

Unlike the change in the abstraction level, there was no clear difference between the Zachman columns with or without the SIM. Both data (what?) and function (how?) were discussed in all sessions at some level, suggesting that these topics are the basis for service identification.

International Journal of Software Engineering & Applications (IJSEA), Vol.6, No.2, March 2015Table 3. Categorization of the service identification sessions

| Session category | Session characteristics and requirements towards SIM |
|---|---|
| Functional service requirement documentation session | Documenting the well-thought business concept and related functional requirements for services.<br><br>R1: To provide a systematic approach to facilitate the documentation of the requirements. |

Discussion topics without and with SIM

| Without SIM | Data | Func | Net | Peop | Time | Mot | With SIM | Data | Func | Net | Peop | Time | Mot |
|---|---|---|---|---|---|---|---|---|---|---|---|---|---|
| Contextual | 3 | 2 | 2 | 2 | 1 | 1 | Contextual | 2 | 2 | 1 | 1 | 1 | 2 |
| Conceptual | 1 | 3 | 1 | 1 | 0 | 3 | Conceptual | 3 | 3 | 1 | 3 | 2 | 2 |
| Logical | 1 | 0 | 1 | 0 | 0 | 0 | Logical | 3 | 3 | 2 | 0 | 2 | 0 |
| Physical | 0 | 0 | 0 | 0 | 0 | 0 | Physical | 0 | 0 | 0 | 0 | 0 | 0 |
| Detailed | 0 | 0 | 0 | 0 | 0 | 0 | Detailed | 0 | 0 | 0 | 0 | 0 | 0 |

Sample size = 3    Sample size = 3

| Session category | Session characteristics and requirements towards SIM |
|---|---|
| Business/SOA alignment session | Two-way communication needed between business stakeholders with conflicting requirements and/or with technical stakeholders understanding the technical feasibility of the requirements.<br><br>R1: To provide a systematic approach to facilitate the discussion.<br>R2: To provide a common ground between technical and business stakeholders, using concepts understandable by both. |

Discussion topics without and with SIM

| Without SIM | Data | Func | Net | Peop | Time | Mot | With SIM | Data | Func | Net | Peop | Time | Mot |
|---|---|---|---|---|---|---|---|---|---|---|---|---|---|
| Contextual | 6 | 6 | 1 | 6 | 1 | 1 | Contextual | 1 | 3 | 0 | 1 | 0 | 0 |
| Conceptual | 4 | 4 | 1 | 3 | 3 | 4 | Conceptual | 5 | 4 | 3 | 3 | 0 | 1 |
| Logical | 0 | 2 | 1 | 1 | 1 | 2 | Logical | 5 | 5 | 5 | 2 | 1 | 4 |
| Physical | 1 | 1 | 0 | 0 | 0 | 0 | Physical | 1 | 1 | 1 | 0 | 0 | 1 |
| Detailed | 2 | 3 | 0 | 1 | 0 | 0 | Detailed | 0 | 0 | 0 | 0 | 0 | 0 |

Sample size = 7    Sample size = 5

## 5. DISCUSSION

This study suggests that the use of a service identification method can make service identification sessions more systematic and steer the discussion towards a more detailed, technical level. The IUC method was found to be a simple enough tool to be explained during a session. Also, the spontaneous reactions of the three project managers and one IT architect that adopted IUCs suggest that IUCs are useful for practitioners.

Gu and Lago [18] have identified the need for an SIM to support the interaction and collaboration of multiple organizations, but found many of the evaluated SIMs to neglect this requirement. All of the studied sessions had participants from several organizations, often from external suppliers, making this requirement first priority for companies similar to the ones studied.

Service identification sessions can be seen as one form of requirement engineering (RE). Generic RE methods can be applied in service identification sessions as well. For example, Burnay et al.

40

[29] propose the Elicitation Topic Map (ETM) to list the topics of discuss and their relative importance when preparing for an RE session. This kind of a topic list could be used as a checklist in service identification, but it should be adapted with topics important for service elicitation.

Another main requirement for an SIM emerged in the sessions when service identification required business stakeholders to discuss and agree on requirements or when technical capabilities limited the feasibility of business requirements. Similar results have been reported in a case study of one telecom operator [7], where the development of a reusable business concept covering both the business process and the IT services was seen as a major breakthrough. In general, better facilitation of IT/business communication has been identified as one of the key enablers for a successful SOA adoption by both Baskerville et al. [6] and Joachim at al. [8].

The validity of ethnographic research is difficult to evaluate as there is no direct way to judge the validity. Maxwell [30] has presented a topology of threats to validity in qualitative research, including descriptive validity, interpretive validity, theoretical validity, generalizability and evaluative validity [30].

Descriptive validity refers to the accuracy of the data, i.e. recording the events correctly and accurately reflecting what was said or happened. Interpretive validity concerns the researcher's capability to correctly interpret what people meant with their statements and behavior [30]. The use of a predefined template and an external framework as a reference to document the sessions can be considered an improvement in both the descriptive, but also the interpretive validity. An external framework enabled the interpretations of the discussions to be made already during the session or immediately after it, meaning the discussion was still fresh in the mind. But still, the assessment of the discussion topics is highly interpretive, and therefore it should be considered as an approximation.

Theoretical validity refers to the researcher's concepts and theorized relationships in the context of the phenomena [30]. In this study, qualitative research has been used to identify the challenges of service identification and the effects of using one kind of an SIM. The main researcher's impression of the challenges was rather similar from session to session, with or without the researcher's direct participation to the discussion. However, making an action mid-session when starting to use the SIM required the researcher to also take a more active role, which may have affected the course of these sessions.

The number of reported sessions and involved companies is too low to generalize the results beyond very similar companies to those studied, especially without the concept of a "typical company," as the requirements on an SIM are likely to differ between different kinds of companies.

Furthermore, as the IUC method is based on use cases, it is vulnerable to the same critique as the use cases in general. Especially, it is the inadequate means to model exception handling that can be seen as an issue for integration use cases. Similarly the lack of means to model long-range dependencies or states between use cases concerns also IUCs, even when the services should be stateless by their nature. Another clear limitation was also identified; the use of video conferencing tools makes it difficult to use IUC to facilitate the discussion.

## 6. CONCLUSIONS

This paper provided an ethnographical look at the challenges in service identification in enterprises, providing a novel approach to tackle the challenges of service identification. The main identified requirement for an SIM was the capability to systematically facilitate service elicitation sessions. The second requirement was the capability to provide a common ground between heterogeneous participants of the sessions in cases where the business requirements and/or technical constraints was not clear. These requirements should be taken into account when selecting suitable SIM for companies similar to the ones studied.

Furthermore, the use of an SIM deepened the discussion during the service identification sessions, allowing the technical constraints to be discussed in business context. Facilitation of IT/business communication affects the success of SOA adoption and can enable service-oriented design of the enterprise.